\documentclass[aps,prd,twocolumn,tightenlines,superscriptaddress,showpacs,byrevtex,amsmath,amssymb,nofloatfix]{revtex4-1}

\usepackage{graphicx}
\usepackage{dcolumn}
\usepackage{bm}
\usepackage{rotating}
\usepackage{bigstrut,bigdelim,multirow}
\usepackage{float}
\usepackage{epstopdf}
\usepackage{mathrsfs}
\usepackage{booktabs}
\usepackage{times}
\usepackage{amsmath}
\usepackage{color}
\usepackage{lineno}
\usepackage{hyperref}
\def\Journal#1#2#3#4{{#1} {\bf #2}, #3 (#4)}
\def\PRL{Phys. Rev. Lett.}
\def\FPC{Front. Phys. China}
\def\RMP{Rev. Mod. Phys.}
\def\EPJC{Eur. Phys. J. C}
\def\PRD{Phys. Rev. D}
\def\PL{Phys. Lett.}
\def\JP{J. Phys.: Conf. Ser.}
\def\NIM{Nucl. Instrum. Methods A}

\def\CPC{Comput. Phys. Commun.}
\def\ChPC{Chin. Phys. C}

\begin{document}
\title{Search for \boldmath$h_{c}\rightarrow\pi^+\pi^-J/\psi$ via $\psi(3686)\rightarrow \pi^0\pi^+\pi^-J/\psi$}

\author{
 \begin{small}
 \begin{center}
M.~Ablikim$^{1}$, M.~N.~Achasov$^{9,d}$, S. ~Ahmed$^{14}$, X.~C.~Ai$^{1}$, M.~Albrecht$^{4}$, D.~J.~Ambrose$^{45}$, A.~Amoroso$^{50A,50C}$, F.~F.~An$^{1}$, Q.~An$^{38,47}$, J.~Z.~Bai$^{1}$
% ...
\\
\vspace{0.2cm}
(BESIII Collaboration)\\
\vspace{0.2cm} {\it
$^{1}$ Institute of High Energy Physics, Beijing 100049, People's Republic of China\\
% ...
$^{4}$ Bochum Ruhr-University, D-44780 Bochum, Germany\\
% ...
$^{9}$ G.I. Budker Institute of Nuclear Physics SB RAS (BINP), Novosibirsk 630090, Russia\\
% ...
$^{14}$ Helmholtz Institute Mainz, Johann-Joachim-Becher-Weg 45, D-55099 Mainz, Germany\\
% ...
$^{38}$ State Key Laboratory of Particle Detection and Electronics, Beijing 100049, Hefei 230026, People's Republic of China\\
% ...
$^{45}$ University of Rochester, Rochester, New York 14627, USA\\
% ...
$^{47}$ University of Science and Technology of China, Hefei 230026, People's Republic of China\\
% ...
$^{50}$ (A)University of Turin, I-10125, Turin, Italy; (B)University of Eastern Piedmont, I-15121, Alessandria, Italy; (C)INFN, I-10125, Turin, Italy\\
% ...
$^{d}$ Also at the Novosibirsk State University, Novosibirsk, 630090, Russia\\
% ...
}\end{center}
 \end{small}
}
\noaffiliation

\date{\today}

\begin{abstract}
  Using a data sample of $448.1\times10^6$ $\psi(3686)$ events collected with the BESIII 
  detector operating at the BEPCII, we perform search for the hadronic transition $h_c\rightarrow\pi^+
  \pi^-J/\psi$ via $\psi(3686)\rightarrow\pi^0h_c$. No signals of the transition 
  are observed, and the upper limit on the product branching fraction $\mathcal{B}(\psi(3686)
  \rightarrow\pi^0h_c)\mathcal{B}(h_c\rightarrow\pi^+\pi^-J/\psi)$ at the 90\% confidence 
  level is determined to be $2.0\times10^{-6}$. This is the most stringent upper limit to date.
\end{abstract}

\pacs{13.66.Bc, 14.40.Pq, 13.25.Gv}
\maketitle

\section{Introduction}
Heavy quarkonium~($Q\bar{Q}$) presents
an ideal environment for testing the interplay between perturbative and nonperturbative
Quantum Chromodynamics (QCD)~\cite{QCD}.  Hadronic transitions between the heavy 
$Q\bar{Q}$ states are particularly interesting. A common approach for calculating these 
transitions is the QCD Multipole Expansion~(QCDME)~\cite{MEMK} for gluon emission. 
The calculation depends on experimental inputs and works well for transitions of heavy 
$Q\bar{Q}$ states below open flavor threshold~\cite{MEM}. But some puzzles remain to 
pose challenge to the theory. For example, the measured ratio $\frac{\Gamma({\Upsilon(2S)\rightarrow\eta\Upsilon(1S)})}{\Gamma
(\psi(2S)\rightarrow\eta J/\psi)}$~\cite{qqbar} is much smaller than the theoretical prediction. 
Hence, more experimental measurements for the transition of heavy $Q\bar{Q}$ are 
desirable to constrain and challenge the theory models.  However to date, the only 
well-measured hadronic transitions in the charmonium sector are those for the $\psi(3686)$.

For charmonium states below the $D\bar D$ threshold, the 
hadronic transitions of the spin-singlet P-wave state $h_c(1^1P_1)$ are one of the best places to test the spin-spin interaction 
between heavy quarks~\cite{prodhc}, but they remain the least accessible experimentally 
because the $h_c(1^1P_1)$ can not be produced resonantly in $e^+ e^-$ 
annihilation or from electric-dipole radiative transitions of the $\psi(3686)$. 
Evidence for the $h_c$ state was reported in $p\bar p\rightarrow h_c\rightarrow
\gamma\eta_c$ by E835~\cite{E835} at Fermilab. The first observation of 
the $h_c$ was reported by CLEO in a study of the cascade decay $\psi(3686)
\rightarrow\pi^0h_c,h_c\rightarrow\gamma\eta_c$~\cite{cleoobs}. With large 
statistics, CLEO measured the $h_c$ mass precisely~\cite{cleomass}, and 
presented evidence for multi-pion decay modes~\cite{cleomultipi}, which 
imply that the $h_c$ state has comparable rates for the decay to hadronic final states 
and the radiative transition to the $\eta_c$ state. Furthermore, for the first time 
the BESIII collaboration measured the branching fractions $\mathcal{B}(\psi(3686)\rightarrow
\pi^0h_c)=(8.4\pm1.3\pm1.0)\times10^{-4}$ and $\mathcal{B}(h_c\rightarrow
\gamma\eta_c)=(54.3\pm6.7\pm5.2)\%$~\cite{bespi0hc}, which were confirmed 
by CLEO~\cite{cleopi0hc}.

The $h_c$ is also expected to decay to lower-mass charmonia state through 
hadronic transitions, but this has not been observed yet. In the framework of QCDME, the 
branching fraction of $h_c\rightarrow\pi\pi J/\psi$~(including charged and neutral 
modes) is predicted to be 2\%~\cite{Kuang1988}, while it is predicted to be 0.05\% when neglecting the 
nonlocality in time~\cite{Py1995}. An experimental measurement is desirable to 
distinguish between these calculations. In this paper, we perform a search for
the hadronic transition $h_c\to\pi^+\pi^-J/\psi$ using a data sample consisting of 
$(448.1\pm2.9)$ million $\psi(3686)$ events~\cite{totpsip} collected at a center-of-mass 
energy of 3.686 GeV, corresponding to the peak of $\psi(3686)$ resonance. 
Considering kinematic limitation and parity conservation, the angular momentum 
between the two-pion system (in a relative S-wave) and $J/\psi$ should be P-wave, and the transition 
rate of $h_c\rightarrow\pi^+\pi^-J/\psi$ is suppressed. Thus, statistical limitation 
and low detection efficiency for the soft pions are the two major challenges to study 
$h_c\rightarrow\pi^+\pi^-J/\psi$. Taking into account the theoretically predicted 
branching fraction for transition $h_c\rightarrow\pi^+\pi^-J/\psi$, the other related 
decay branching fractions from Particle Data Group (PDG)~\cite{PDG2016} 
and the total number of $\psi(3686)$ used in this analysis and without consideration 
of detection efficiency, the signal yield of $\psi(3686)\rightarrow\pi^0h_c,\pi^0
\rightarrow\gamma\gamma, h_c\rightarrow\pi^+\pi^-J/\psi,J/\psi\rightarrow l^+l^-
~(l=e,\mu)$ is excepted to be 600 and 15 for the predictions of Refs.~\cite{Kuang1988} and \cite{Py1995}, 
respectively.

This paper is structured as follows: in Section II the BESIII
detector is described and details of the Monte Carlo (MC) samples 
are given. In Section III, the analysis strategy, event selection criteria
and background analysis are introduced. Section IV presents the
estimation of the upper limit, and Section V provides the systematic
uncertainties of the measurement. Finally, a short summary and a
discussion of the result are given in Section VI.

\section{BESIII DETECTOR AND MONTE CARLO SIMULATION}

The BESIII detector is designed to facilitate physics research in the 
$\tau$-charm region in $e^+e^-$ annihilations with center-of-mass energies from 2 to 4.6\,GeV at 
the Beijing Electron Positron Collider II (BEPCII). The detector has a 
geometric acceptance of 93\% of the solid angle and mainly consists 
of five components: (1) a helium-gas-based main drift chamber (MDC) 
for tracking and particle identification using the specific energy loss $\mathrm{d}E/\mathrm{d}x$. 
The expected charged particle momentum resolution at 1\,GeV and 
$\mathrm{d}E/\mathrm{d}x$ resolution are 0.5\% and 6\%, respectively. 
(2) a plastic scintillator time-of-flight system with an intrinsic time 
resolution of 80\,ps in the barrel region and 110\,ps in the end-cap region. 
(3) a CsI(Tl) crystal calorimeter (EMC) with an energy resolution better 
than $2.5\%$ in the barrel region and $5\%$ in the end-cap region, 
and a position resolution better than 6\,mm for 1\,GeV electrons and 
photons. (4) a superconductive solenoid magnet with a central field of 
1.0\,Tesla. (5) a muon chamber system composed of nine barrel layers 
and eight end-cap layers of resistive plate chambers with a spatial 
resolution better than 2\,cm. More details on the construction and 
capabilities of BESIII detector may be found in Ref.~\cite{bes3design}.

The optimization of event selection criteria, study of backgrounds 
and determination of detection efficiency are based on samples of MC simulated events. 
A GEANT4-based~\cite{geant4} software is used to describe the 
geometry of the BESIII detector and simulate the detector response. 
A MC sample of 506 million generic $\psi(3686)$ decays (`inclusive MC sample') is 
generated to study the background processes. The $\psi(3686)$
resonance is generated by KKMC~\cite{KKMC} with final state
radiation (FSR) effects handled with PHOTOS~\cite{PHOTOS}. The
known decay modes are generated by EvtGen~\cite{EvtGen} with branching
fractions set to the world average values according to the PDG~\cite{PDG2014}; 
the remaining unknown charmonium decays are generated with LundCharm~\cite{LCharm}.
The signal channel $\psi(3686)\rightarrow\pi^0 h_c,h_c
\rightarrow\pi^+\pi^-J/\psi$ is excluded from the inclusive sample.

\begin{figure*}[htbp]
\centering
\includegraphics[scale=0.41]{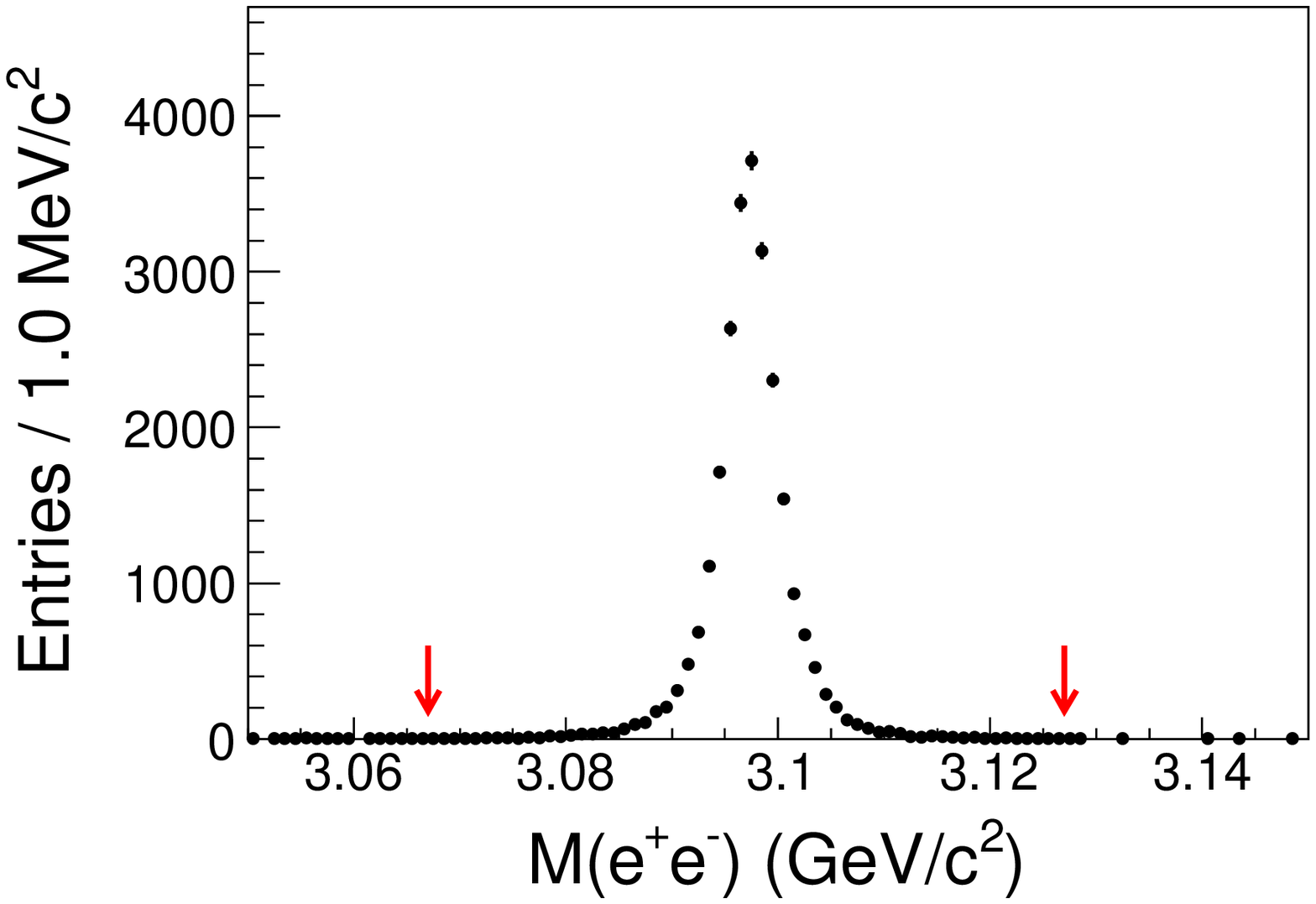}
\hspace{0.1cm}
\includegraphics[scale=0.41]{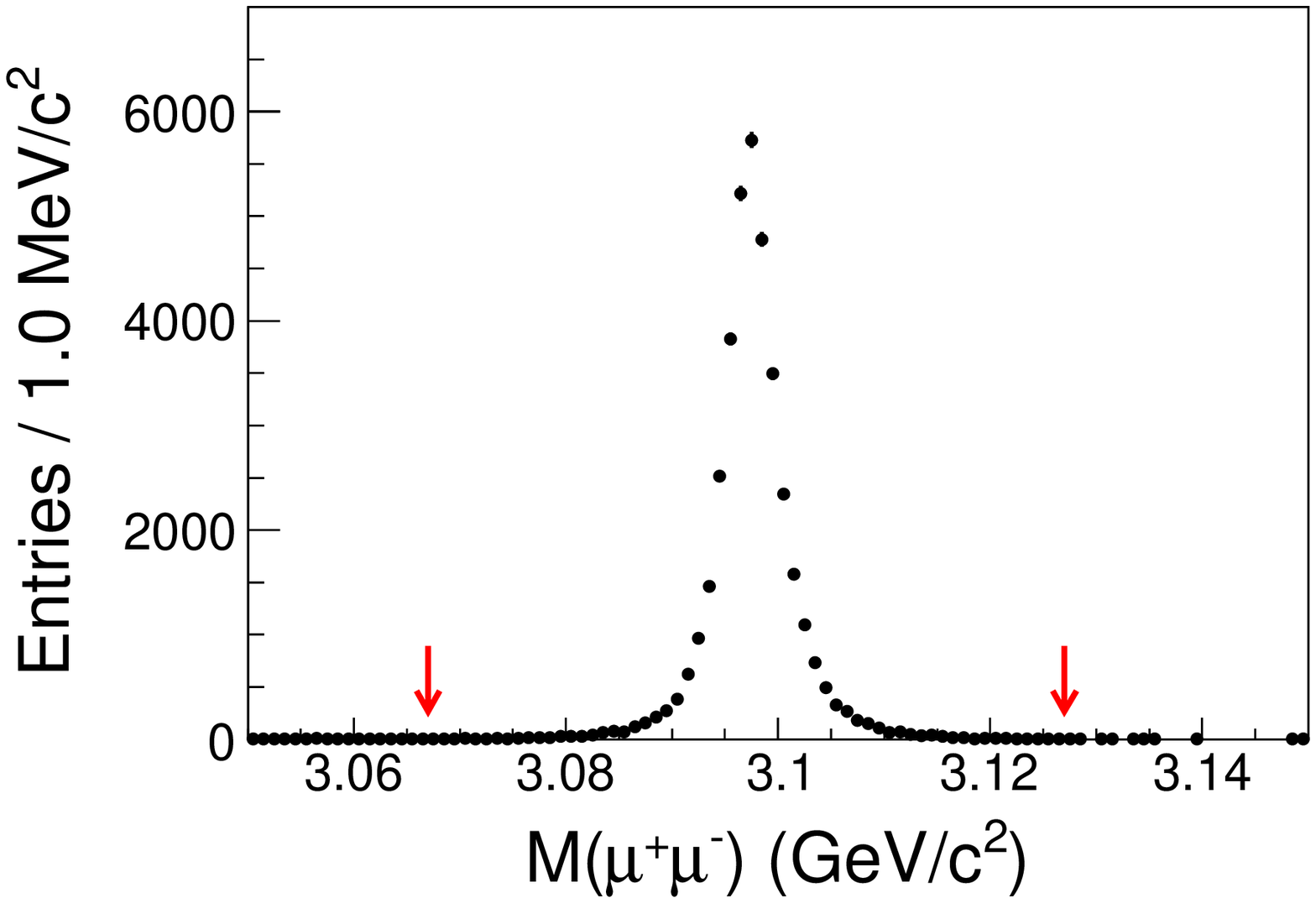}
\caption{Distributions of $M(e^+ e^-)$ (left) and $M(\mu^{+}\mu^{-})$
  (right) in data. The arrows show the signal region.}
\label{Mll}
\end{figure*}

The signal MC sample of $\psi(3686)\rightarrow\pi^0h_c, h_c\rightarrow
\pi^+\pi^-J/\psi$ is generated uniformly in phase space with the $\pi^0$ 
decaying to two photons and the $J/\psi$ decaying to $l^+l^-~(l=e,\mu)$. 
The MC sample of $\psi(3686)\rightarrow\eta J/\psi$ with $\eta$ decaying 
to $\pi^0\pi^+\pi^-$ and $J/\psi$ decaying to $l^+l^-$ is generated to study 
the background and determine the detection efficiency of this process. 
The angular distribution of the $\eta$ is modeled as $1+\cos^2\theta_\eta$, 
where $\theta_\eta$ is the angle between $\eta$ momentum and the positron 
beam in the rest frame of $\psi(3686)$. The decay $\eta\rightarrow\pi^0
\pi^+\pi^-$ is generated by EvtGen~\cite{EvtGen} with the measured Dalitz plot
amplitude~\cite{eta_dalitz}, and $\pi^0\rightarrow\gamma\gamma$ by 
a phase space distribution. The $J/\psi$ decays to $l^+l^-$ are generated 
with an angular distribution of $1+\cos^{2}\theta_{l}$, where $\theta_{l}$ 
is the angle between the $l^+$ momentum in the $J/\psi$ rest frame and the 
$J/\psi$ momentum in the $\psi(3686)$ rest frame.

\section{METHODOLOGY AND EVENT SELECTION}
A relative measurement strategy is used to measure $h_c\rightarrow\pi^+\pi^-
J/\psi$ according to
\begin{eqnarray}
&&\mathcal{B}(\psi(3686)\rightarrow\pi^0h_c)\mathcal{B}(h_c\rightarrow\pi^+\pi^-J/\psi) = \nonumber \\ &&\frac{N_{\mathrm{sig}}^{\mathrm{obs}}}{N_{\mathrm{ref}}^{\mathrm{obs}}}\frac{\epsilon_{\mathrm{ref}}}{\epsilon_{\mathrm{sig}}}
\mathcal{B}(\psi(3686)\rightarrow\eta J/\psi)\mathcal{B}(\eta\rightarrow\pi^0\pi^+\pi^-).
\label{eqn}
\end{eqnarray}
The decay $\psi(3686)\rightarrow\pi^0h_c\rightarrow\pi^0\pi^+\pi^-J/\psi$ is 
the signal mode, and the decay $\psi(3686)\rightarrow\eta J/\psi\rightarrow
\pi^0\pi^+\pi^-J/\psi$, which has the same final state as the signal, serves as the 
reference mode. These two processes will be selected simultaneously. Then the product
$\mathcal{B}(\psi(3686)\rightarrow\pi^0h_c)\mathcal{B}(h_c\rightarrow\pi^+\pi^-J/\psi)$ 
can be obtained by the ratio of the numbers of observed events 
$N_{\mathrm{sig}}^{\mathrm{obs}}/N_{\mathrm{ref}}^{\mathrm{obs}}$ and the 
ratio of detection efficiencies $\epsilon_{\mathrm{ref}}/\epsilon_{\mathrm{sig}}$
of these two processes. With this relative measurement method, most of the 
systematic uncertainties in the efficiencies and that of the total number of $\psi(3686)$ 
events cancel.

Charged tracks are reconstructed from hits in the MDC and are required to 
originate from the interaction point, \emph{i.e.} passing within 10\,cm to the interaction 
point in the beam direction and 1\,cm in the plane perpendicular to the beam. 
In addition, the polar angle $\theta$ of each track is required to satisfy $|\cos\theta|<0.93$. 
Electromagnetic showers are reconstructed from clusters in the EMC. A good 
photon candidate is an isolated shower that is required to have energy larger than 25\,MeV in the barrel 
region of the EMC~($|\cos\theta|<0.8$) or 50\,MeV in the end-cap regions~
($0.86<|\cos\theta|<0.92$). Showers in the transition region between the barrel and the 
end-cap are removed since they are not well reconstructed. In addition, timing
information from the EMC ($0\leq t\leq700$\,ns) is used to suppress electronic 
noise and energy deposits unrelated to the event. 

For events of interest, including $\psi(3686)\rightarrow\pi^0h_c, h_c\to \pi^+\pi^-
J/\psi$ (signal mode), and $\psi(3686)\rightarrow\eta J/\psi, \eta\to\pi^0\pi^+\pi^-$ 
(reference mode), we require that there are four good charged tracks with zero 
net charge and at least two good photon candidates. The track momentum is used to separate 
leptons and pions since the momenta of leptons from $J/\psi$ decay
are higher than $1$\,GeV$/c$. Charged tracks with momenta less than $1$\,GeV$/c$ 
are assumed to be pions, while the remaining two tracks are taken as leptons. Electrons 
and muons are identified according to the ratio of energy ($E$) deposited in the EMC 
and momentum ($p$) measured in MDC. Tracks with $E/pc>0.7$ are taken as 
electrons, and those with $E/pc<0.3$ are identified as muons. A pair of pions 
with opposite charge and a pair of leptons with same flavor and opposite charge are required.
Photon pairs with invariant mass in the region $120<M(\gamma\gamma)<145$\,MeV$/c^2$ 
are combined into $\pi^0$ candidates. To avoid bias in choosing the best combination, 
all combinations due to multiple $\pi^{0}$ candidates are retained. Only 0.5\% of all events
contain more than one $\pi^0$ candidate, and this is modeled well in the simulation.
The $\pi^+\pi^-$ invariant mass $M(\pi^+\pi^-)$ should be larger than 0.3\,GeV/$c^2$ 
to reject backgrounds from $\pi^0\pi^0J/\psi$ with $\gamma$ converting into an $e^+e^-$ 
pair in the beam pipe or inner wall of the MDC.

\begin{figure}[htbp]
\centering
\includegraphics[scale=0.43]{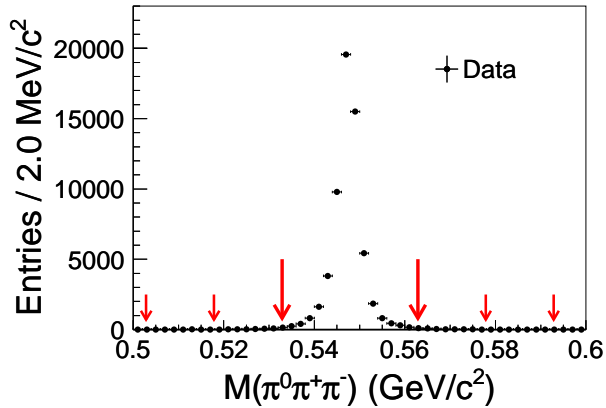}
\caption{Distribution of $M(\pi^0\pi^+\pi^-)$ of data,
  the longer red arrows indicate the signal region of $\psi(3686)\rightarrow\eta
  J/\psi$ and the shorter red arrows correspond to the sideband regions. }
\label{fiteta}
\end{figure}

\begin{figure*}[!htbp]
\centering
\includegraphics[scale=0.41]{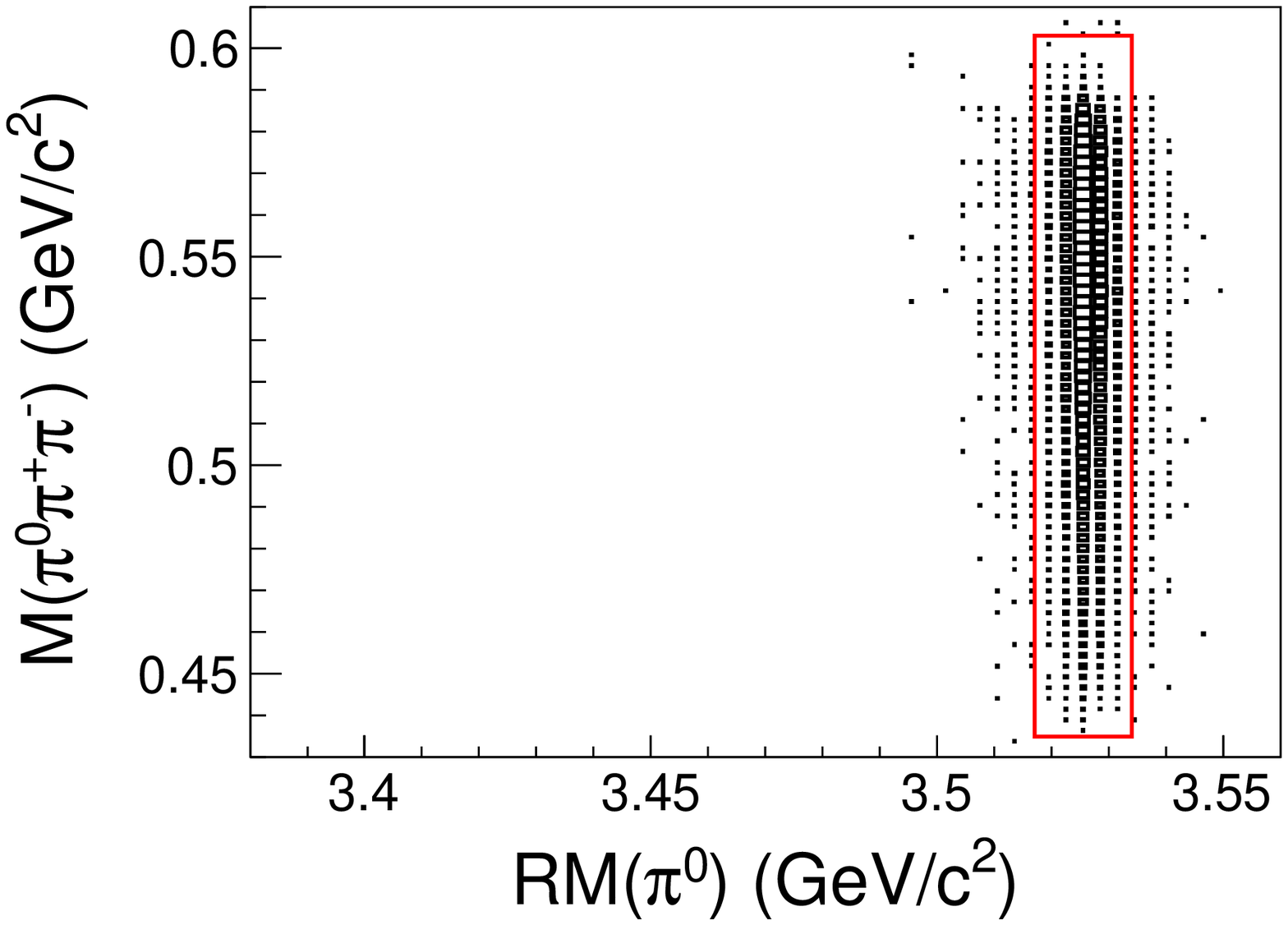}
\hspace{0.2cm}
\includegraphics[scale=0.41]{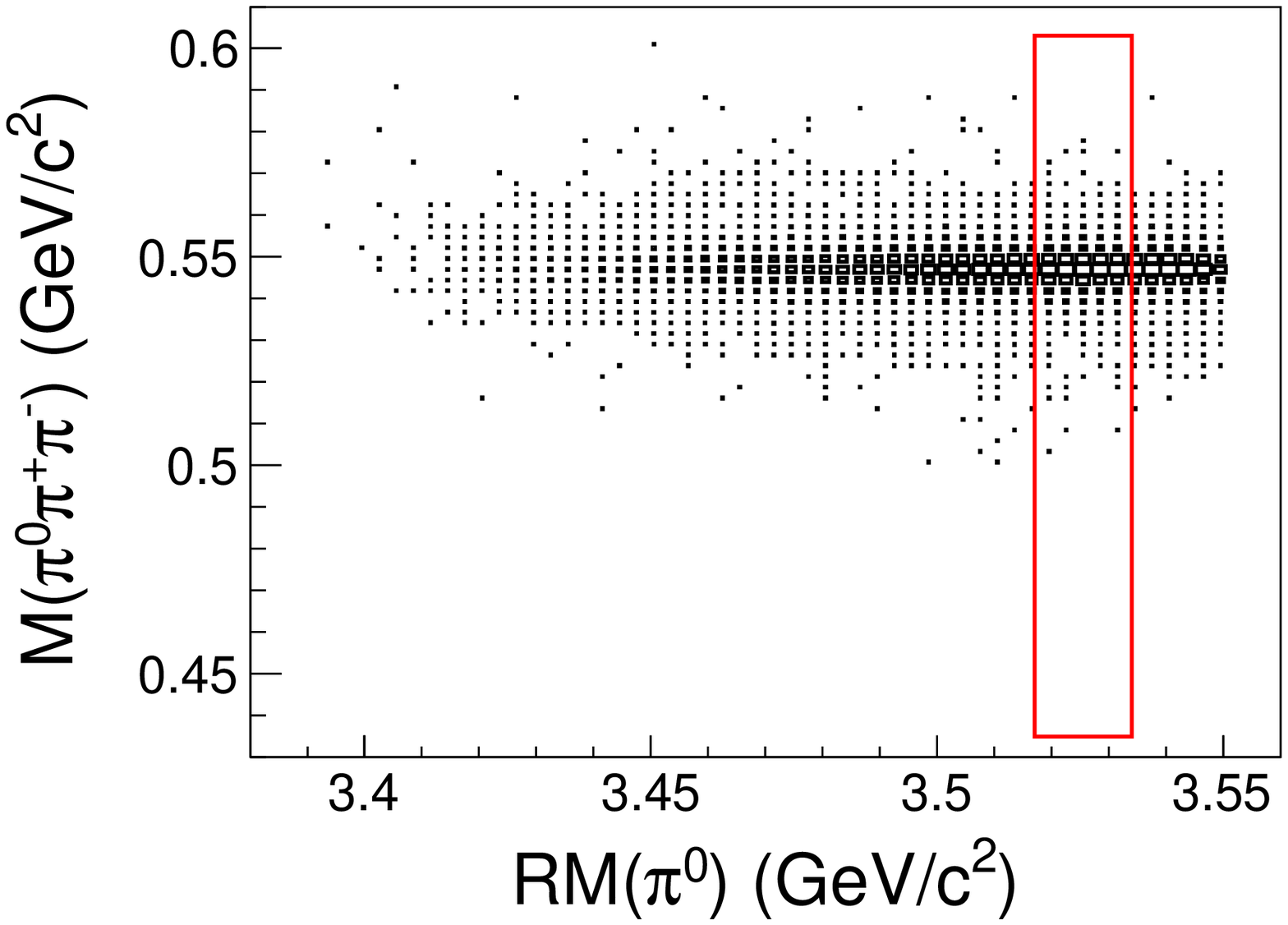}
\caption{Two-dimensional distributions of $M(\pi^0\pi^+\pi^-)$ versus $RM(\pi^0)$
        for the signal MC sample (left) and data (right). The red box indicates the $h_c$ signal 
        region.}
\label{scat}
\end{figure*}

A five-constraint (5C) kinematic fit is performed for the $\pi^0\pi^+\pi^-l^+l^-$
combination enforcing energy and momentum conservation and constraining
the invariant mass of the photon pair to the $\pi^0$ nominal mass~\cite{PDG2016}. 
Events with $\chi_{\mathrm{5C}}^2<60$ are accepted for further analysis. 
After imposing these criteria, clear $J/\psi$ peaks with low background levels
are observed in both the $e^+e^-$ and $\mu^+\mu^-$ invariant mass distributions, 
as shown in Fig.~\ref{Mll}. For the selection of $J/\psi$ candidates, the invariant 
mass of lepton pairs $M(l^+l^-)$ is required to be in the $J/\psi$ mass region, \emph{i.e.\ }
$|M(l^+l^-)-M(J/\psi)|<30\,$MeV$/c^2$, where $M(J/\psi)$ is the nominal mass of 
the $J/\psi$~\cite{PDG2016}.

Based on studies of the inclusive MC sample, the dominant surviving event 
candidates are from $\psi(3686)\rightarrow\eta J/\psi,\eta\rightarrow\pi^0\pi^+\pi^-$, 
while background from events with different final states is negligible. A clear $\eta$ peak 
with a low level of background is observed in the $\pi^0\pi^+\pi^-$ invariant mass 
spectrum, $M(\pi^0\pi^+\pi^-)$, as shown in Fig.~\ref{fiteta}.

In order to validate the event selection criteria, we calculate the branching fraction 
$\mathcal{B}(\psi(3686)\rightarrow\eta J/\psi)$ and compare it with a previous BESIII
measurement~\cite{besetaJpsi}, where $\eta$ is reconstructed via two photons 
and only the first set of the data sample of $(107.0\pm0.8)$ million $\psi(3686)$ taken 
in 2009~\cite{totpsip} was used. In our calculation, the yield of $\psi(3686)\rightarrow\eta
J/\psi,\eta\rightarrow\pi^0\pi^+\pi^-$ is obtained by counting events in the $\eta$ signal 
region and subtracting the events in the $\eta$ sideband region. The $\eta$ signal region is 
defined as $|M(\pi^0\pi^+\pi^-)-M(\eta)|<15\,$MeV$/c^2$, where $M(\eta)$ is the $\eta$ 
nominal mass~\cite{PDG2016}. It covers about 99.2\% of the $\eta J/\psi$ signal 
according to the MC simulation. The $\eta$ sideband region is defined as $30<|M(\pi^0\pi^+\pi^-)
-M(\eta)|<45\,$MeV$/c^2$. Using the same sample of $107$ million $\psi(3686)$ events, we 
obtain $\mathcal{B}(\psi(3686)\rightarrow\eta J/\psi)=(33.89\pm0.27(\mathrm{stat.}))\times10^{-3}$, 
which is consistent with the previous measurement $(33.75\pm0.17(\mathrm{stat.})\pm0.86(\mathrm{syst.}))\times10^{-3}$.

\section{UPPER LIMIT on \boldmath $\mathcal{B}(\psi(3686)\to\pi^0h_c)\mathcal{B}(h_c\to\pi^+\pi^-J/\psi)$}
\begin{figure}[htbp]
\centering
\includegraphics[scale=0.42]{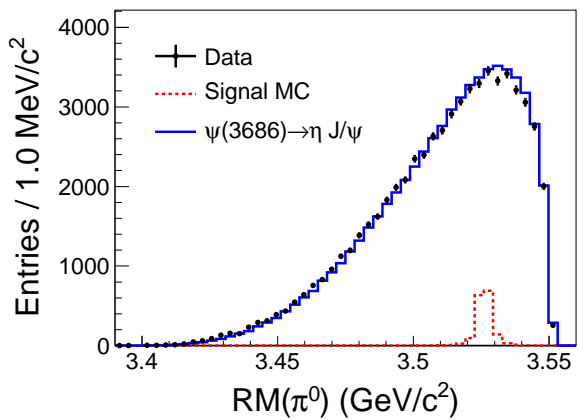}
\caption{Distribution of $RM(\pi^0)$ after the 1C kinematic fit. Black
  dots with error bars show data. The red dashed histogram shows the 
  MC simulated signal shape (with arbitrary normalization). The blue 
  solid histogram is the MC distribution of the reference mode.}
\label{project}
\end{figure}

The two-dimensional distributions of $M(\pi^{0}\pi^{+}\pi^{-})$ versus the $\pi^0$ 
recoil mass~$RM(\pi^0)$ for the signal MC sample and data are shown in Fig.~\ref{scat}, 
and the distribution of $RM(\pi^0)$ is shown in Fig.~\ref{project}. To improve the 
resolution, $RM(\pi^0)$ is calculated using the four-momenta after constraining
the invariant mass of the photon pair to the $\pi^0$ nominal mass~\cite{PDG2016}~(1C). 
The process $\psi(3686)\rightarrow\eta J/\psi$ is clearly dominant, but no obvious 
signal events from $\psi(3686)\rightarrow\pi^0h_c,h_c\rightarrow\pi^+\pi^-J/\psi$ are 
observed.

\begin{table}[htp]
  {\caption{Summary table.~In order:~upper limit on the number of observed signal events
        $(N_\mathrm{sig}^\mathrm{obs})^\mathrm{up}$, the number of observed background 
        events $N_\mathrm{bkg}^\mathrm{obs}$, signal efficiency ($\epsilon_\mathrm{sig}$),
        the number of observed events of reference mode ($N_\mathrm{ref}^\mathrm{obs}$),
        efficiency of reference mode ($\epsilon_\mathrm{ref}$), statistical uncertainty
        ($\sigma^{\mathrm{stat}}$) and total uncertainty ($\sigma^{\mathrm{tot}}$)}
    \label{tb-I}}
  \begin{tabular}{p{3.2cm}p{2cm}<{\centering}}
    \hline
    \hline
    Quantity                                                    &  Value                            \\
    \hline
    $(N^{\mathrm{obs}}_{\mathrm{sig}})^\mathrm{up}$     &  2.44           \\
    $N^{\mathrm{obs}}_{\mathrm{bkg}}$       & 0                                     \\
    $\epsilon_{\mathrm{sig}}$                        &  2.52\%                           \\
    $N^{\mathrm{obs}}_{\mathrm{ref}}$        &  $31611\pm178$             \\
    $\epsilon_{\mathrm{ref}}$                        &  8.25\%                           \\
    $\sigma^{\mathrm{stat}}$                         &  0.57\%                          \\
    $\sigma^{\mathrm{tot}}$                           &  15.4\%                          \\
    \hline
    \hline
  \end{tabular}
\end{table}

In order to obtain the yield of the decay of interest, we veto $\psi(3686)\rightarrow\eta
J/\psi$ by imposing the further requirement $|M(\pi^0\pi^+\pi^-)-M(\eta)|>32$\,MeV/$c^2$. 
For $\psi(3686)\rightarrow\eta J/\psi$, events off the $\eta$ peak region are those with 
bad resolution and large $\chi^{2}_{\mathrm{5C}}$. Thus, to further suppress the events 
from $\psi(3686)\to\eta J/\psi$ which are far from the $\eta$ signal region, a tighter requirement 
$\chi_{5c}^2<15$ is imposed. With the above requirements, 99.99\% of the $\psi(3686)\to\eta
J/\psi$ backgrounds are removed according to MC simulation. No events in data survive 
in the full region of $RM(\pi^{0})$. Based on a study of the inclusive MC sample, there 
are only two background events from $\psi(3686)\rightarrow2(\pi^{+}\pi^{-})\pi^{0}$ left. 
Neither event is in the signal region of the $h_c$, which is defined as $3.517<RM(\pi^{0})<3.534$\,GeV/$c^2$. 
We therefore take the expected number of observed background events $\bar{N}_{\mathrm{bkg}}^{\mathrm{obs}}$ 
in the signal region as zero. The upper limit on the number of observed signal events $N^{\mathrm{obs}}_{\mathrm{sig}}$ 
at the 90\% confidence level (C.L.) is 2.44, which is estimated by using the Feldman-Cousins 
frequentist approach~\cite{s-signal-yeild} without considering the systematic uncertainties. All 
the numbers used to extract the upper limit of signal yield are summarized in Table~\ref{tb-I}. It 
is assumed that $N^{\mathrm{obs}}_{\mathrm{sig}}$ and $N^{\mathrm{obs}}_{\mathrm{bkg}}$ 
follow Poisson distributions.
The number of events and the efficiency of the reference
mode are obtained with the same method and requirements as in Section III, only with
$\chi^{2}_{\mathrm{5C}}<15$ instead of $\chi^{2}_{\mathrm{5C}}<60$.

The upper limit on the product branching fraction 
$\mathcal{B}(\psi(3686)\rightarrow\pi^0h_c)\mathcal{B}(h_c\rightarrow\pi^+\pi^-J/\psi)$ 
at the 90\% C.L. is obtained by replacing $N^{\mathrm{obs}}_{\mathrm{sig}}$ in 
Eq.~(\ref{eqn}) with $(N^{\mathrm{obs}}_{\mathrm{sig}})^\mathrm{up}(1+
(N^{\mathrm{obs}}_{\mathrm{sig}})^\mathrm{up}(\sigma^{\mathrm{tot}})^2/2)$ 
using the method proposed by Cousins and Highland~\cite{ch} to incorporate 
the systematic uncertainty. The branching fractions of $\psi(3686)\to\eta J/\psi$ 
and $\eta\rightarrow\pi^0\pi^{+}\pi^{-}$ are taken from PDG~\cite{PDG2016}. 
The upper limit on $\mathcal{B}(\psi(3686)\rightarrow\pi^0h_c)\mathcal{B}(h_c
\rightarrow\pi^+\pi^-J/\psi)$ at the 90\% C.L. is found to be $2.0\times10^{-6}$.

\section{SYSTEMATIC UNCERTAINTY}
In this analysis, the upper limit is obtained with a relative measurement
strategy defined by Eq.~(\ref{eqn}). Since the signal mode and reference 
mode have same final states, and the uncertainty associated with the 
detection efficiency, \emph{i.e.}~trigger, photon detection, tracking and PID for 
charged tracks, $\pi^0$ reconstruction, and the 5C kinematic fit cancel. The systematic 
uncertainty due to the $M(\pi^0\pi^+\pi^-)$ resolution is less than 0.1\% 
and is negligible.

\begin{table}[htp]
  {\caption{Summary of systematic uncertainties}
  \label{tb-II}}
  \begin{tabular}{p{3.3cm}cp{3.3cm}}
    \hline
    \hline
    Sources                                                                   & Systematic uncertainties (\%)    \\
    \hline
    $\mathcal{B}(\psi(3686)\rightarrow\eta J/\psi)$       & 1.5              \\
    $\mathcal{B}(\eta\rightarrow\pi^{0}\pi^{+}\pi^{-})$   & 1.2              \\
    MC model                                                               & 15.2              \\
    \hline
    Total                                                                        & 15.4             \\
    \hline
    \hline
  \end{tabular}
\end{table}

The $M(\pi^+\pi^-)$ spectrum in the final state of $h_c\to\pi^+\pi^-J/\psi$ 
is unclear due to its unknown dynamics. In the nominal analysis, the signal 
MC sample is generated uniformly in the phase space without considering 
the angular distribution. In order to estimate the related uncertainties of the MC 
model, an alternative signal MC sample is generated by assuming a pure 
P-wave production between the two-pion system (S-wave) and $J/\psi$, 
where the production amplitude is proportional to the third power of the 
momentum of the $\pi^+\pi^-$ system. The difference in detection efficiency 
between the two MC samples, 15.2\%, is taken as the systematic uncertainty 
associated with the MC model.

The branching fractions of $\psi(3686)\rightarrow\eta J/\psi$ and
$\eta\rightarrow\pi^0\pi^+\pi^-$ are taken from the PDG~\cite{PDG2016}.
The uncertainties of the branching fractions, 1.5\% and 1.2\%, are 
considered as systematic uncertainties. The individual systematic 
uncertainties are summarized in Table~\ref{tb-II}.  Assuming that all sources 
of systematic uncertainties are independent, a total systematic 
uncertainty of 15.4\% is obtained by taking the quadratic sum of the individual 
contributions.

\section{SUMMARY}
In summary, a search for the hadronic transition $h_c\rightarrow\pi^{+}\pi^{-}J/\psi$ 
is carried out via $\psi(3686)\rightarrow\pi^0h_c, h_c\rightarrow \pi^{+}\pi^{-}J/\psi$.
No signal is observed.  The upper limit of the product of branching fractions
$\mathcal{B}(\psi(3686)\rightarrow\pi^0h_c)\mathcal{B}(h_c\rightarrow\pi^{+}\pi^{-}J/\psi)$
at the 90\% C.L. is determined to be $2.0\times10^{-6}$. Using the PDG
value for the branching fraction of $\psi(3686)\rightarrow\pi^0h_c$ of
$(8.6\pm1.3)\times10^{-4}$~\cite{PDG2016}, the upper limit on
$\mathcal{B}(h_c\rightarrow\pi^+\pi^-J/\psi)$ is determined to be
$2.4\times10^{-3}$, which is the most stringent upper limit to date. Neglecting the
small phase space difference between the charged and neutral $\pi\pi$ modes and assuming isospin symmetry, 
we obtain $\mathcal{B}(h_c\rightarrow\pi\pi J/\psi)<3.6\times10^{-3}$~(including 
charged and neutral modes) at the 90\% C.L. 
It is noted that the measured branching fraction is smaller than the prediction in 
Ref.~\cite{Kuang1988} by one order in magnitude, but does not contradict that 
in Ref.~\cite{Py1995}.

\section*{Acknowledgement}

The BESIII collaboration thanks the staff of BEPCII and the IHEP computing center for their strong support. This work is supported in part by National Key Basic Research Program of China under Contract No. 2015CB856700; National Natural Science Foundation of China (NSFC) under Contracts Nos. 11205117, 11235011, 11322544, 11335008, 11425524, 11575133; the Chinese Academy of Sciences (CAS) Large-Scale Scientific Facility Program; the CAS Center for Excellence in Particle Physics (CCEPP); the Collaborative Innovation Center for Particles and Interactions (CICPI); Joint Large-Scale Scientific Facility Funds of the NSFC and CAS under Contracts Nos. U1232201, U1332201; CAS under Contracts Nos. KJCX2-YW-N29, KJCX2-YW-N45; 100 Talents Program of CAS; National 1000 Talents Program of China; INPAC and Shanghai Key Laboratory for Particle Physics and Cosmology; Hubei Nuclear Solid Physics Key Laboratory; German Research Foundation DFG under Contracts Nos. Collaborative Research Center CRC 1044, FOR 2359; Istituto Nazionale di Fisica Nucleare, Italy; Joint Large-Scale Scientific Facility Funds of the NSFC and CAS	under Contract No. U1532257; Joint Large-Scale Scientific Facility Funds of the NSFC and CAS under Contract No. U1532258; Koninklijke Nederlandse Akademie van Wetenschappen (KNAW) under Contract No. 530-4CDP03; Ministry of Development of Turkey under Contract No. DPT2006K-120470; National Science and Technology fund; The Swedish Resarch Council; U. S. Department of Energy under Contracts Nos. DE-FG02-05ER41374, DE-SC-0010504, DE-SC0012069; U.S. National Science Foundation; University of Groningen (RuG) and the Helmholtzzentrum fuer Schwerionenforschung GmbH (GSI), Darmstadt; WCU Program of National Research Foundation of Korea under Contract No. R32-2008-000-10155-0.

\end{document}